# Large CISS Polarization from the Hierarchy Between Transport and Geometrical Spin Currents

Zeev Vager
Weizmann Institute of Science, Rehovot, Israel

Email zeev.vager@weizmann.ac.il

Abstract

Large spin polarization observed in chiral-induced spin selectivity (CISS) remains difficult to explain quantitatively. Experimental polarizations measured in chiral molecular systems are often substantially larger than expected from weak microscopic spin–orbit interaction in organic materials. We revisit the geometrical spin current introduced previously for electrons constrained to curved paths and propose a scale argument relevant to transport through double-stranded DNA (dsDNA). Estimation of the intrinsic geometrical spin-current scale using representative dsDNA parameters yields current magnitudes substantially larger than measured transport currents in many CISS experiments. This hierarchy allows transport to be interpreted as weak leakage from underlying curved spin-current states, in which the geometrical spin current is compensated by an opposing orbital current. Within this picture, large CISS polarization emerges through transport selection of pre-existing spin-current branches.

## 1. Introduction

Chiral molecular systems frequently exhibit substantial spin polarization during electron transport. Double-stranded DNA (dsDNA) remains among the most studied examples.

Recent reviews emphasize that despite extensive theoretical progress, quantitative explanation of the large observed CISS magnitudes remains an open problem [4,5].

A previous work by Dan Vager and Zeev Vager introduced geometrical spin–orbit current for electrons constrained to curved paths and discussed spin order in curved molecular systems [1].

The present Letter does not modify that framework. Instead, we introduce a quantitative scale argument that was not considered in Ref. [1].

The missing element in Ref. [1] was not the existence of geometrical spin current but its comparison with experimentally relevant transport currents.

To our knowledge, the role of the intrinsic geometrical spin-current scale $I_S$ relative to measured transport currents $I_{meas}$ has not been examined.

Recent work has emphasized that internal molecular current itself may play a more central role in CISS than previously assumed and may contribute even in the absence of intrinsic spin–orbit coupling [6]. The present work follows a different route and proposes that the relevant quantity controlling large CISS polarization is the hierarchy between intrinsic geometrical spin current and measured transport current.

Rather than invoking an additional amplification mechanism, large CISS polarization is proposed to arise from a hierarchy of current scales.

Specifically, when

$$I_{meas} \ll |I_S| \qquad (1)$$

transport is weak leakage from underlying bound states carrying geometrical spin current.

Within this picture, large observed spin polarization reflects selection of pre-existing spin-current branches.

The experimentally observed importance of spin–orbit coupling in CISS is not disputed here. The present Letter addresses a different question: whether the hierarchy between intrinsic geometrical spin currents and measured transport currents can account for the large observed spin polarization.

## 2. Geometrical spin current

For electrons constrained to move along a curved path, the effective current contains orbital and spin-induced contributions [1]:

$$J = J_O + J_S \qquad (2)$$

For the two spin branches:

$$J_S^+ = -\alpha\rho_+/m_e \qquad (3)$$

$$J_S^- = +\alpha\rho_-/m_e \qquad (4)$$

Where $\alpha$ is the geometrical spin–orbit interaction parameter, $\rho$ is the local transport-active electronic density, $m_e$ is the electron mass.

The present interpretation takes ρ to represent one-dimensional transport-active electronic density within the transport energy window rather than total molecular electronic density.

## 3. Estimate for dsDNA

Assume transport involves a substantial fraction of one effective transport-active electron per repeat unit:

$$\rho_{tr} \approx f/d \qquad (5)$$

with $0.1 \leq f \leq 1$.

For dsDNA, the repeat spacing $d \approx 0.34$ nm, together with the helix radius, determines the curvature $\kappa \approx 0.76\ \text{nm}^{-1}$. Since the geometrical momentum scale is $\alpha = (\hbar/2)\kappa$, this gives

$$\alpha/m_e \approx 4 \times 10^4\ \text{m s}^{-1}. \qquad (6)$$

The characteristic geometrical spin-current scale is then

$$I_S = e\alpha\rho_{\text{tr}}/m_e, \qquad (7)$$

This gives

$$I_S \approx 2 - 20\ \mu\text{A} \qquad (8)$$

for $0.1 \leq f \leq 1$.

Many reported CISS transport currents in molecular and dsDNA experiments are considerably smaller than this estimated geometrical spin-current scale and frequently lie in the low-current molecular transport regime, especially in [2,3] where photoelectrons were detected. Therefore, the weak-leakage condition

$$I_{meas} \ll I_S \qquad (9)$$

is assumed.

4. Microscopic leakage mechanism

Consider collective electronic states on a fixed-curvature molecular path. In the bound-state limit, electrons carry geometrical spin current $J_S$, compensated by an opposite orbital contribution $J_O$, such that

$$J = J_O + J_S = 0 \qquad (10)$$

Selection of one spin branch, such as Eq. (3), simultaneously selects a definite direction of the compensating orbital current. Weak coupling to external transport channels then allows partial leakage of this current. Because the leaking carriers originate from the selected spin-current branch, they inherit its spin polarization.

In this picture, weak measured transport current does not generate polarization but reveals pre-existing spin-selected current states.

The present Letter does not derive the microscopic leakage dynamics but proposes that the hierarchy of current scales provides a useful organizing principle for interpreting large CISS polarization.

## 5. Experimental consequences

The present interpretation suggests:

(i) spin polarization should correlate with $I_{meas}/I_S$;

(ii) increasing transport current should reduce polarization once $I_{meas}/I_S$ approaches unity;

(iii) increasing curvature should increase $I_S$;

(iv) reversal of current direction should reverse selected spin polarization.

## 6. Conclusions

The intrinsic geometrical spin current estimated for dsDNA appears substantially larger than the measured transport currents in many CISS experiments. This hierarchy suggests that the observed transport may reveal pre-existing spin-selected current states through weak leakage, thereby providing a possible explanation for the large spin polarizations observed in CISS. A three-dimensional extension of the geometrical spin-current formalism, incorporating both curvature and torsion, will be presented elsewhere.

**Fig. 1.** Conceptual dependence of CISS polarization on the ratio $I_{\mathrm{meas}}/I_S$, where $I_S$ is the intrinsic geometrical spin-current scale. Large polarization is expected only in the weak-leakage regime $I_{\mathrm{meas}} \ll I_S$. As the measured transport current approaches the intrinsic spin-current scale, the weak-leakage condition is lost, and polarization rapidly decreases.

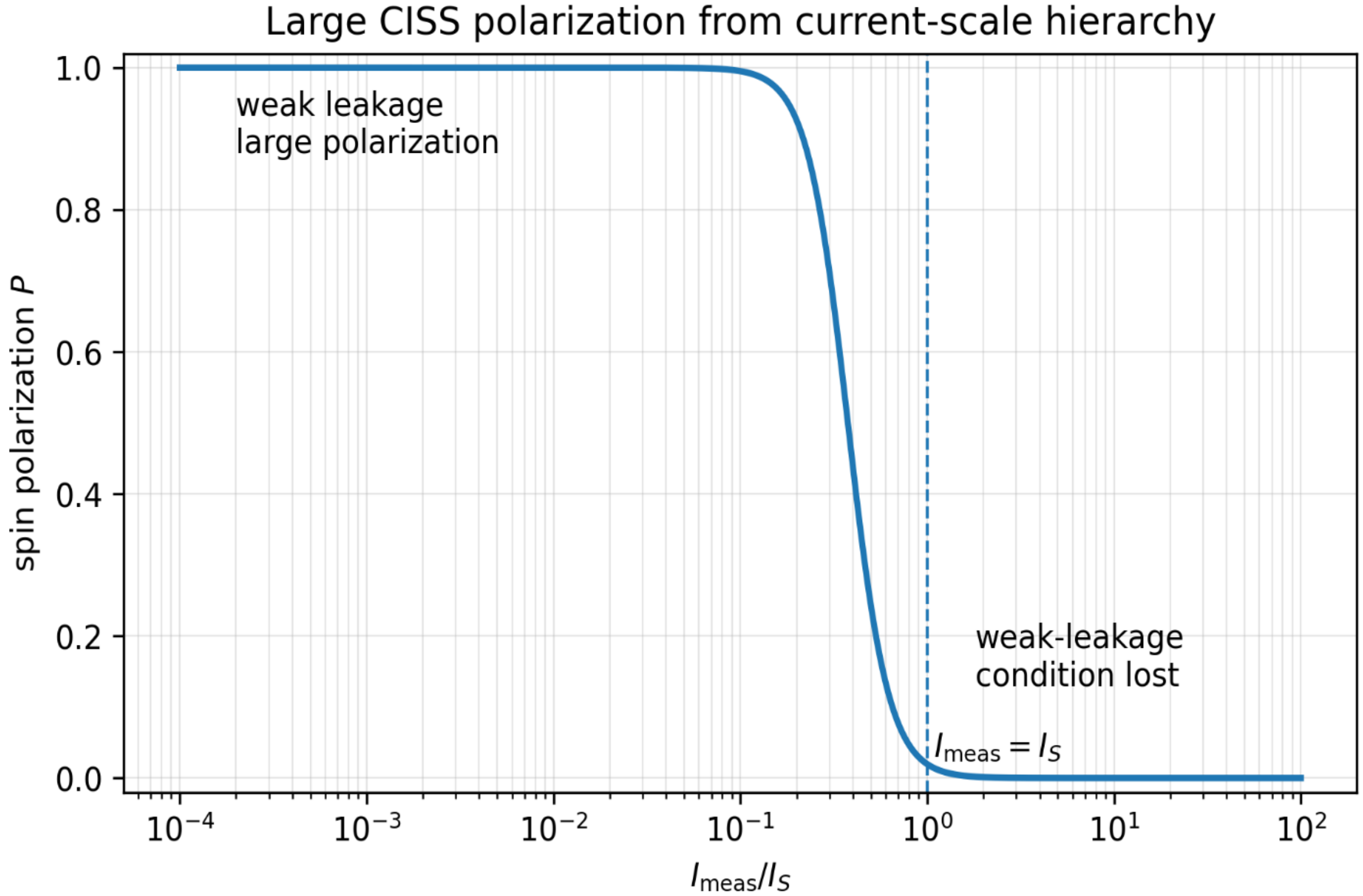